\documentclass[twocolumn,showpacs,preprintnumbers,amsmath,amssymb]{revtex4-1}

\usepackage{amssymb,amsmath,latexsym}
\usepackage{dcolumn} 

\usepackage{epsfig,psfrag,subfigure}
\usepackage{graphicx,psfrag,subfigure}
\usepackage[usenames,dvipsnames]{color}
\usepackage{bbold}

\usepackage{bm}
\usepackage{epsfig,amsopn}
\usepackage{graphicx}
\usepackage{amsmath,amssymb}
\usepackage{amsthm}
\usepackage{enumerate}
\usepackage{multirow}

\newcommand\bea{\begin{eqnarray}}
\newcommand\eea{\end{eqnarray}}
\newcommand\beq{\begin{equation}}
\newcommand\eeq{\end{equation}}
\newcommand\bsq{\begin{subequations}}
\newcommand\esq{\end{subequations}}

\newcommand{\noi}{\noindent}
\newcommand{\bib}{\bibitem}

\def\nn{\nonumber}
\def\f{\frac}
\def\al{\alpha}
\def\om{\omega}
\def\bom{{\bar \omega}}

\def\de{\delta}
\def\ep{\epsilon}

\def\si{\sigma}
\def\Do{\partial}
\def\mc{\mathcal}
\def\vr{\varrho}
\def\la{\langle}
\def\ra{\rangle}
\def\mbb{\mathbb}
\def\ua{\uparrow}
\def\dg{\dagger}
\def\De{\Delta}
\def\da{\downarrow}
\definecolor{dgreen}{rgb}{0.30,0.36,0.10}
\definecolor{ddgreen}{rgb}{0.10,0.3,0.10}
\definecolor{crimson}{rgb}{0.85,0.16,0.20}
\definecolor{dred}{rgb}{0.9,0.35,0.5}
\definecolor{dblue}{rgb}{0.1,0.1,0.7}

\begin{document}

\title{Model of resistances in systems of Tomonaga-Luttinger liquid wires}
\author{Abhiram Soori and Diptiman Sen}
\affiliation{Centre for High Energy Physics, Indian Institute of Science, 
Bangalore 560012, India}
\date{\today}

\begin{abstract}
In a recent paper, we combined the technique of bosonization with the 
concept of a Rayleigh dissipation function to develop a model for resistances
in one-dimensional systems of interacting spinless electrons 
[arXiv:1011.5058]. 
We also studied the conductance of a system of three 
wires by using a current splitting matrix $M$ at the junction.
In this paper we extend our earlier work in several ways. The power dissipated
in a three-wire system is calculated as a function of $M$ and the voltages 
applied in the leads. By combining two junctions of three wires, we examine a 
system consisting of two parallel resistances. We study the conductance of 
this system as a function of the $M$ matrices and the two resistances; we find
that the total resistance is generally quite different from what one expects 
for a classical system of parallel resistances. We will do a sum over paths 
to compute the conductance of this system when one of the two resistances 
is taken to be infinitely large. Finally we study the conductance of a 
three-wire system of interacting spin-1/2 electrons, and show that the 
charge and spin conductances can generally be different from each other.

\end{abstract}

\maketitle

\section{Introduction}

For non-interacting electrons, the conductance of a 
ballistic quantum wire is well-known to be quantized in units of $2e^2/h$ 
at low temperatures~\cite{buttiker,datta,wees}. This remains valid when 
interactions between the electrons in the wire are taken into account, 
provided that there are no sources of backscattering (such as impurities
and junctions) and that the wire is connected to leads where there are no 
interactions~\cite{maslov,pono,safi1,safi2,thomale}. Thus, if 
the wire is modeled as a Tomonaga-Luttinger liquid (TLL) and the interaction 
strength is given by the Luttinger parameter $K_W$, the conductance of a clean
wire does not depend on $K_W$. This breaks down if there are impurities in a 
wire with interacting electrons. In that case, the impurity strengths vary 
with the length scale according to some renormalization group (RG) equations, 
and the conductance depends on $K_W$ and other parameters like the length of
the wire, the distances between the impurities, and the 
temperature~\cite{kane-furusaki,expt}. A considerable amount of work has 
also been done on junctions of several quantum wires both 
theoretically~\cite{sand,nayak,lal,chen,chamon,meden,das,giuliano,bella1,agar} 
and experimentally~\cite{fuhrer,terrones}. In these systems, the conductance 
matrix again becomes length scale dependent due to the interactions.
A junction of three quantum wires with interacting spin-1/2 electrons 
has been studied in Ref.~\cite{hou}, and it has been found that some of the 
fixed points of the RG equations exhibit different charge and spin
conductances. Thus impurities in a single wire or a junction of three or
more wires effectively give rise to a resistance which leads to power 
dissipation. There have been some studies of power dissipation on the edges 
of a quantum Hall system~\cite{wen} and at a junction of quantum 
wires~\cite{bella2}. However, there has been relatively little 
discussion in the literature of the effects of an extended region of 
dissipation (a resistive patch) within the framework of TLL theory or 
bosonization which is the most efficient way to study the effects
of interactions~\cite{boson}. 

To remedy this situation, we recently
introduced a formalism which can combine the technique of bosonization
with the classical notion of resistance; both a single wire and a system of 
three wires with a junction were studied using this formalism~\cite{soori}. 
Our analysis was restricted to spinless electrons and zero temperature. In 
contrast to this, Ref.~\cite{rech} considered the effect of an extended 
region of inhomogeneity in a quantum wire at low temperatures; it was shown 
that this leads to weak backscattering which gives rise to a resistance 
which is linear in the temperature. 

It is useful to briefly
recapitulate our earlier work~\cite{soori}. We introduced the resistances 
phenomenologically using a Rayleigh dissipative function~\cite{gold}. Our 
treatment was classical in the sense that the resistance was taken to be 
purely a source of power dissipation, and the microscopic quantum mechanical 
origins of the resistance were not specified. This is equivalent to treating 
scattering by the resistance as a phase incoherent process; a consequence of 
this is that the resistances of different patches add up in series with 
no effects of interference. We found expressions for the conductance
of a single wire and of a three-wire system with a junction (this was
described by a current splitting matrix $M$ which is orthogonal) in terms of 
the resistances in the wires. The conductance can be calculated even when the 
Luttinger parameter $K_W$, the velocity $v_W$ of the quasiparticles and 
the resistivity $r$ all vary with the spatial coordinate $x$ in an arbitrary 
way in the different wires. Remarkably, we found that the conductance of the
three-wire system is independent of the Luttinger parameter $K_W$ (which is 
determined by the strength of the interaction between the electrons near the 
junction) if the matrix $M$ is invariant under time reversal; this remains 
true no matter what the values of the resistances are in the three wires. 

In this paper, we will extend the ideas introduced in Ref.~\cite{soori}
in several ways. For the sake of completeness, we will briefly present our 
earlier results for a single wire in Sec. II and a three-wire system in Sec. 
III. We will then examine the issue of power dissipation in a three-wire 
system in Sec. IV. The dissipation occurs due to the resistances in the wires
and the contact resistance in the leads. We will study the dependence of the 
power dissipation on the junction matrix $M$
and the relative magnitudes of the voltages applied in the leads of the system.
In Sec. V, we will study a system of two resistances in parallel; the system 
consists of two junctions of three wires. We will make a detailed study of
the total conductance in terms of the two junction matrices $M$ and the two 
resistances; the conductance will be calculated using a scattering approach.
We find that the classical expression for the effective resistance of two 
parallel resistances is recovered only in one special case. In Sec. VI, we 
will carry out a sum over paths to calculate the conductance of the same 
system in the special case that one of the two resistances is infinitely 
large. In Sec. VII, we will generalize our results to the case of spin-1/2 
electrons. We will argue there that the charge and spin conductances can 
generally be different from each other. We will summarize our results in
Sec. VIII.

\section{Single wire}
\label{single-wire}

\begin{figure}[htb]
\begin{center}
\epsfig{figure=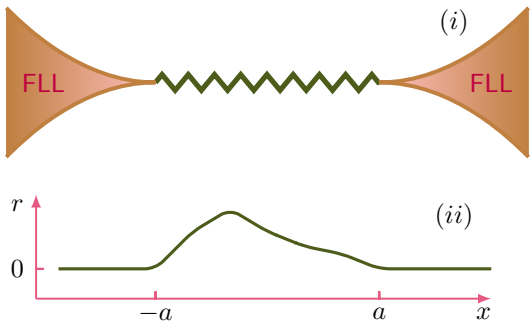,width=7cm}
\end{center}
\caption{(i) Schematic picture of a resistive region connected to Fermi 
liquid leads (FLL). (ii) A graph showing the resistivity as a function 
of the length; $\int_{-a}^a dx~r(x) = R$.} \label{ResLeads} \end{figure}

\subsection{Equation of motion}

We start with the analysis of a single wire containing interacting spinless 
electrons. In the absence of any impurities, the bosonic Lagrangian for the 
system is given by
\beq L ~=~ \int_{-\infty}^\infty dx ~[\f{1}{2vK} (\Do_t \phi)^2 ~-~ \f{v}{2K}
(\Do_x \phi)^2], \label{lag1} \eeq
where $K$ and $v$ respectively denote the Luttinger parameter and velocity of
the bosonic quasiparticles; we will allow $K$ and $v$ to vary with $x$ within 
a finite region which we will take to be $-L/2 < x < L/2$. (For noninteracting
electrons, $K=1$, while for short-range repulsive interactions like a 
screened Coulomb interaction, we have $K<1$). The limit of noninteracting 
electrons, with $K=1$ and $v=v_F$ (the Fermi velocity), is used to model the 
two- or three-dimensional Fermi liquid leads situated in the regions $|x| > 
L/2$. In the leads, the frequency and wave number of a plane wave are related 
by $\om = v_F |k|$. The electron charge density $n$ and current $j$ can be 
expressed in terms of the bosonic field by the relations: $n = -e \Do_x 
\phi/\sqrt{\pi}$ and $j = e \Do_t \phi/ \sqrt{\pi}$, where $e$ is the 
electron charge; these satisfy the equation of continuity 
$\Do_t n + \Do_x j = 0$. To describe resistances phenomenologically, 
we introduce a Rayleigh dissipation function
\beq {\cal F} ~=~ \f{1}{2} ~\int_{-\infty}^\infty dx ~r ~j^2, 
\label{diss} \eeq
where the resistivity $r$ can vary with $x$. We then obtain the 
following equation of motion as described in Ref.~\cite{soori}
\beq \f{1}{vK} ~\Do_t^2 \phi ~-~ \Do_x \left(\f{v}{K} \Do_x \phi \right) ~
+~ \f{e^2}{\pi} ~r ~\Do_t \phi ~=~ 0. \label{eom} \eeq
Note that we have set $\hbar =1$, so that $e^2/(2\pi) = e^2/h$.

\subsection{Scattering approach}\label{single-wire-scatt-app}

We will now derive an expression for the DC conductance for a general 
resistance profile $r(x)$, and will show that only the total resistance 
$R = \int dx r(x)$ of all the resistive patches appears in the final 
expression for $\si_{dc}$. Also, we will show that a $\de$-function 
resistance with the same integrated value of $R$ gives the same value 
of the conductance. 

As mentioned earlier, we set $K=1$ and $v=v_F$ in the leads. However, $K(x)$ 
and $v(x)$ can have any profile in the wire region given by $|x|<a$. 
Similarly, we will assume that the resistivity $r(x) =0$ for $|x|>a$, but 
can have any profile in the region $|x|<a$ such that $\int_{-a}^a dx~r(x) = 
R$ [one such profile is illustrated in Fig.~\ref{ResLeads} (ii)]. As 
described in Ref.~\cite{soori}, in the scattering approach, 
a plane wave with frequency $\om$ is incident on the resistive patch, the 
reflection and transmission amplitudes are calculated as functions of $\om$,
and finally the limit $\om~\to~0^+$ is taken to obtain the expression for 
$\sigma_{dc}$. For a plane wave incident from left with $k=\om /v_F$, the 
spatial part of the solution $\phi_k (x,t) = f_k (x) e^{-i\om t}$ outside 
the resistive patch is given by
\bea f_k &=& e^{ikx} + s_k ~e^{-ikx} ~~~{\rm for}~~~ x \le -a, \nn \\
&=& t_k ~e^{ikx} ~~~{\rm for}~~~ a \le x. \label{sktk} \eea
Up to zero-th order in $\om$ (and hence in $k$) the solution of Eq.~\eqref{eom}
is $\phi=f_k=c$, where $c$ is a constant. Therefore we can see from 
Eq.~\eqref{sktk} that 
\beq t_k ~=~ 1 ~+~ s_k ~=~ c {\rm~~~up~to~zero-th~order~in}~\om.
\label{s_eq1} \eeq
Since we are eventually interested in the limit $\om \to 0^+$, we can work out
the solution to Eq.~\eqref{eom} up to first order in $\om$ and then take the 
limit $\om \to 0^+$. We first rewrite Eq.~\eqref{eom} up to first order in 
$\om$, 
\beq -~ \Do_x \left(\f{v}{K} \Do_x f_k \right) ~-~ i \om \f{e^2}{\pi} ~r ~
f_k ~=~ 0. \label{eom2} \eeq
In the second term in Eq.~\eqref{eom2}, we can replace $f_k$ by a constant $c$ 
(= $t_k$ up to zero-th order in $\om$) since this term has a factor of $\om$ 
already. Integrating this equation from $x=-a-\ep$ to $a+\ep$, we obtain 
$ik v_F [t_k - (1-s_k)] = - i \om t_k (e^2 R/\pi)$ which is the same as
\beq \Big(1 + \f{e^2 R}{\pi}\Big) t_k + s_k ~=~ 1. \label{s_eq2} \eeq
Solving Eqs.~\eqref{s_eq1}~and~\eqref{s_eq2}, we obtain
$t_k = 1/[1 +\f{e^2R}{2\pi}]$. Note that this expression for $t_k$ is 
correct only in the limit $\om \to 0^+$ (or $k \to 0^+$). Hence, 
\bea \si_{dc} &=& \f{e^2}{2\pi} ~t_{k \to 0^+} \nn \\ 
&=& \f{e^2}{2\pi} ~\f{1}{1 +\f{e^2R}{2\pi}}. \label{dc-cond} \eea

One can easily redo the calculations from Eq.~\eqref{sktk} to 
Eq.~\eqref{dc-cond}
and see that the result remains unchanged if we choose a $\de$-function 
resistivity profile described by $r(x) = R \de (x-x_0)$, where $|x_0|<a$.
We will use this fact later in Sec.~\ref{pulse}. We thus see that $\si_{dc}$
does not depend on the precise functional forms of $K(x)$, $v(x)$ and $r(x)$, 
as long as $K$ and $v$ are equal to the constants 1 and $v_F$ in the leads,
and $R$ is the total resistance of the wire. An implication of this is 
that the effective resistance of two or more resistive patches in series will
be given by the sum of the individual resistances. 

In this context, it is worthwhile to look at the phase coherent and phase
incoherent transport in the literature (see for instance pp.~125-129 of 
Ref.~\cite{datta}). In general, the effective resistance of two resistances
can have an extra term that depends on some phase factors at the two resistive
patches in addition to the sum of the individual resistances. However, the 
effective resistance reduces to the sum of individual resistances in the 
incoherent limit. In this sense, our formalism assumes that the
system is phase incoherent.

\section{Three wires with a junction}
\label{three-wires}

\begin{figure}[htb]
\begin{center} 
\epsfig{figure=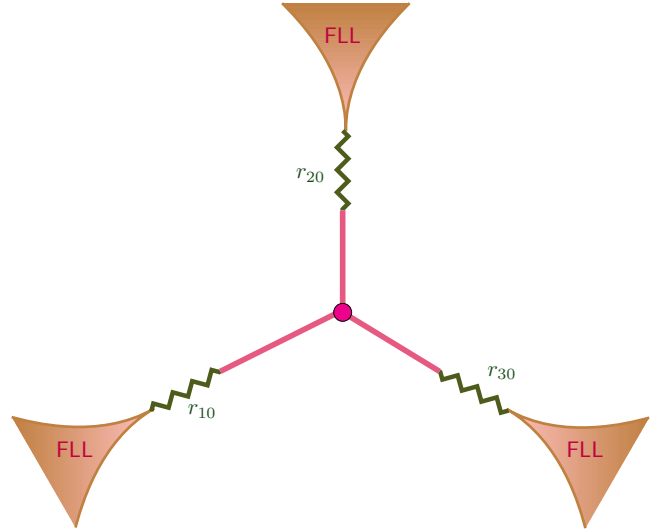,width=8.5cm}
\end{center}
\caption{Schematic diagram of a three-wire junction with interacting 
regions close to the junction (pink solid region), dissipative regions (green 
zig-zag region further from the junction), and Fermi liquid leads (brown 
shaded region furthest from the junction).} \label{Y-jn-diagram} \end{figure}

Let us consider a junction of three TLL wires as shown in 
Fig.~\ref{Y-jn-diagram}. We will assume that each wire has three regions: 

\noi (i) $0\le x_i \le L_{i1} \ne 0 $ --- the wire region around the junction 
where $K(x_i)=K_W$ and $v(x_i) = v_W$; elsewhere $K(x_i)= 1$ and $v(x_i)=v_F$,

\noi (ii) $L_{i1} \le x_i \le L_{i2}$ --- the dissipative region where 
$r (x_i)= r_{i0}$; elsewhere $r (x_i)=0$, and

\noi (iii) $x_i \ge L_{i2}$ --- the semi-infinite leads.

Here $i$ labels the wires, and on wire $i$, the coordinate $x_i$ runs from 
$0$ to $\infty$, with $x_i =0$ denoting the junction point. The regions 
$x_i \ge L_{i2}$ model the two- or three-dimensional leads which are 
assumed to be Fermi liquids with no interactions between the electrons; 
we therefore set $K=1$ and $v=v_F$ in those regions.

\subsection{Green's function approach}

We now follow Ref.~\cite{maslov} and write 
\bea I_i &=& \sum_{j=1}^{3} \int_0^{L_{j2}} dx_j' \int_{-\infty}^\infty 
\f{d\om}{2\pi} e^{-i\om t} \si_{ij,\om} (x_i,x_j') E_{\om}(x_j'), \nn \\
\label{kubo} \eea
in the linear response regime, where $E_{\om}(x_j')$ is the Fourier component 
of the electric field $E(x_j',t)$ on wire $j$, and $\si_{ij,\om} (x_i,x_j')$ 
is the nonlocal ac conductance matrix. Using the Kubo formula, we then obtain 
\beq \si_{ij,\om} (x_i,x_j') ~=~ -\f{e^2 \bom}{\pi} ~{\cal G}_{ij,\bom}(x_i,
x_j'), \label{sigma-G} \eeq
where $\bom = - i \om$, and
\beq {\cal G}_{ij,\bom} (x_i,x_j') ~=~ \int_0^\infty \f{d\tau}{2\pi} \la 
T^*_{\tau} \phi_i (x_i,\tau) \phi_j (x_j',0) \ra e^{i\bom \tau} \label{G} \eeq
is the Fourier transform of the bosonic field in imaginary time, $\tau = it$.
The expressions for $\tau$ and $\bom$ in terms of $t$ and $\om$ arise as 
follows. The actions in real (Minkowski) time $t$ and imaginary (Euclidean) 
time $\tau$ are given by
\bea S_M &=& \int \int dt dx ~[\f{1}{2vK} (\Do_t \phi)^2 ~-~ \f{v}{2K}
(\Do_x \phi)^2], \nn \\
S_E &=& \int \int d\tau dx ~[\f{1}{2vK} (\Do_\tau \phi)^2 ~+~ \f{v}{2K}
(\Do_x \phi)^2] \eea
respectively. The requirement that the exponentials appearing in a path 
integral formulation be equal to each other, i.e., $e^{iS_M} = e^{-S_E}$, 
implies that $\tau = it$. Secondly, we want an outgoing plane wave in the lead
of wire $i$ to be given by $e^{i(\om x_i/v_F - \om t)}$ in terms of a real 
frequency $\om$ and $e^{-\bom x_i/v_F - i\bom \tau}$ in terms of an imaginary
frequency $\bom$ (this expression for an outgoing wave will be used
in the boundary condition (iii) given below). The two planes will be 
identical if $\bom = - i \om$.

The Green's function satisfies the equation
\bea & & \Big[ -\Do_{x_i} \Big(\f{v(x_i)}{K(x_i)} \Do_{x_i} \Big) + 
\f{\bom^2}{v(x_i)K(x_i)} + \f{e^2 \bom}{\pi} r (x_i) \Big] \nn \\
& & ~~{\cal G}_{ij,\bom} (x_i,x_j') ~=~ \de_{ij} ~\de (x_i-x_j'), 
\label{G-Eq-Mo} \eea 
with the following three boundary conditions:

\noi (i) ${\cal G}_{ij,\bom} (x_i,x_j')$ is continuous at $x_i=x_j'$ (where 
$0<x_j'< L_{j2}$) and $-\f{v(x_i)}{K(x_i)} \Do_{x_i} {\cal G}_{ij,\bom} 
(x_i,x_j')|_{x_j'-\ep}^{x_j'+\ep} = \de_{ij}$,

\noi (ii) ${\cal G}_{ij,\bom}(x_i,x_j')$ and $- \f{v(x_i)}{K(x_i)} \Do_{x_i} 
{\cal G}_{ij, \bom}(x_i,x_j')$ are continuous at $x_i=L_{i1}~{\rm and}~L_{i2}$,

\noi (iii) if ${\cal G}_{ij,\bom}(x_i,x_j') = A_{ij} e^{\bom x_i/v_W} + B_{ij}
e^{-\bom x_i/v_W}$ for $0<x_i<{\rm min}(x_j',L_{i1})~\de_{ij}+L_{i1}(1-
\de_{ij})$, then $B =-M ~A$, where $M$ is the current splitting matrix at the
junction~\cite{sand,nayak,lal,chen,chamon,meden,das,giuliano,bella1,agar}.

\noi The boundary condition in (iii) encodes the fact that the incoming 
and outgoing currents (and hence the bosonic fields) at the junction are 
related by the matrix $M$. Various constraints at the junction such as current
conservation and unitarity of the evolution of the system in real time (i.e., 
no power is dissipated exactly at the junction) imply that each row and column
of $M$ must add up to unity and that $M$ must be orthogonal. It turns out
that for a junction of three wires, the possible $M$ matrices must belong to
one of two classes both of which are parameterized by a single parameter 
$\theta$~\cite{sand,nayak,lal,chen,chamon,meden,das,giuliano,bella1,agar}: 
(a) det$(M_1) = 1$ and (b) det$(M_2) = -1$; these can be expressed as:
\bea M_1 = \left( \begin{array}{ccc} 
a & b & c \\
c & a & b \\
b & c & a \end{array} \right) &{\rm ~and~}&
M_2 = \left( \begin{array}{ccc} 
b & a & c \\
a & c & b \\
c & b & a \end{array} \right),~~~ \label{m1-m2} \eea
where $a = (1+2\cos \theta)/3$, $b = (1-\cos \theta + \sqrt{3} \sin \theta)/
3$, and $c = (1-\cos \theta - \sqrt{3} \sin \theta)/3$. We note that 
$(M_2)^2 = {\mbb 1}$ for any value of $\theta$; this relation will be used 
below.

Note that by introducing the orthogonal matrix $M$, we have made the 
assumption that there is no dissipation exactly at the junction.
The analysis is simpler if the junction, which governs how the incoming 
currents are distributed amongst the different wires, is separated from 
the dissipative regions which lie some distance away from the junction.

Solving Eq.~\eqref{G-Eq-Mo} with the above boundary conditions and 
taking the limit $\bom \to 0^+$, we obtain the following expression for the 
dc conductance matrix: 
\bea G &=& - ~\f{e^2 K_W}{\pi} ~[ {\mbb 1} + M + K_W ({\mbb 1} - M) ({\mbb 1} 
+ \f{e^2}{\pi} \mbb R)]^{-1} \nn \\
& & ~~~~~~~~\times ~[ {\mbb 1} - M], \label{cond} \eea
where $\mbb R$ is a $3 \times 3$ diagonal matrix with ${\mbb R}_{ii} = R_i =
r_{i0} (L_{i2}-L_{i1})$; we note that $R_i$ is simply the total resistance in 
wire $i$. The conductance matrix relates the outgoing current $I_i$ to the 
potential $V_i$ applied in lead $i$ as $I_i = \sum_j G_{ij} V_j$. One can show
in general that each row and column of $G$ must add up to zero; the columns 
adding up to zero is a consequence of current conservation ($\sum_i I_i$ must 
be zero), while the rows must add up to zero because each of the $I_i$ must 
vanish if the $V_j$'s have the same values in all the wires.

Eq.~\eqref{cond} can be understood as a combination of the conductance of a 
system with no resistances ($R_i=0$) and resistances $R_i$ on the 
three wires. Let us denote the conductance with no resistances by
\beq G_0 = - ~\f{e^2 K_W}{\pi} ~[ {\mbb 1} + M + K_W ({\mbb 1} - M)]^{-1} ~[
{\mbb 1} - M]. \label{cond2} \eeq
If $V'_i$ denote the potentials at the points $x_i = L_{i1}$ (i.e., the 
points which lie after the interacting regions but before the dissipative 
regions), we have $I_i = \sum_j G_{0,ij} V'_j$. Further, $I_i = (V'_i - V_i)/
R_i$. Combining these equations with $I_i = \sum_j G_{ij} V_j$, we obtain
\beq G ~=~ [{\mbb 1} ~-~ G_0 {\mbb R}]^{-1} ~G_0. \label{cond3} \eeq
This relation will be used in Sec.~\ref{spin-12} below.

\subsection{Scattering approach}\label{junc-scatt}

Eq.~\eqref{cond} can be derived in general using the equation of 
motion approach in the limit $\om \to 0^+$ in the same way as described 
above for the single wire case. Let $\phi_i (x_i,t) = f_i (x_i) 
e^{-i\om t}$ denote the bosonic field in real time which satisfies 
Eq.~\eqref{eom} in wire $i$. Current conservation at the junction 
implies that $\sum_i \Do_t \phi_i (x_i \to 0+\ep, t) = 0$ which 
implies that $\sum_i f_i (x_i \to 0+\ep) = 0$. [Note that this
is a different condition than the one used in the single wire case 
where $f(x)$ was assumed to be continuous everywhere; for the case 
of more than two wires, it is more convenient to assume $\sum_i f_i 
(x_i \to 0+\ep) = 0$ rather than the continuity of $f_i (x_i \to 0+\ep)$ 
between different values of $i$]. We now assume that
\bea f_i (x_i \to 0+\ep) &=& a_i e^{-ik' x_i} ~+~ b_i e^{ik' x_i}, \nn \\
f_i (x_i \to \infty) &=& \al_i e^{-ik x_i} ~+~ \beta_i e^{ik x_i},
\label{fields} \eea
where $a_i, ~\al_i$ denote the incoming fields and $b_i, ~\beta_i$ 
denote the outgoing fields. Assuming that the Luttinger parameter and 
the velocity are given by $K_W$ and $v_W$ as $x_i \to 0+\ep$ and by 
1 and $v_F$ as $x_i \to \infty$, we must have $v_W k' = v_F k = \om$. 
In the limit $\om \to 0^+$, the field on wire $i$ is given by $a_i + b_i$ 
at $x_i \to 0+\ep$ and by $\al_i + \beta_i$ as $x_i \to \infty$ at 
zero-th order in $\om$, $k$ and $k'$. Since $f_i (x_i)$ equal to 
a constant is a solution of Eq.~\eqref{eom} for $\om = 0$, we must 
have $a_i + b_i = \al_i + \beta_i$. Next, the coefficients $a_i$ 
and $b_i$ are related by the current splitting matrix at the junction, 
$b = - M ~a$. The coefficients $\al_i$ and $\beta_i$ must be related 
by the conductance matrix in the leads, $\beta = - [(2\pi/e^2) G + {\mbb I}]~
\al$; this follows from the statement that $\sum_i (\al_i + \beta_i) 
= \sum_i (a_i + b_i) = 0$. Finally, we integrate Eq.~\eqref{eom} 
from $x_i = 0+\ep$ to $\infty$, ignoring the first term which 
is of order $\om^2$ and setting the third term equal to $-i \om 
(e^2/\pi) (\al_i + \beta_i) {\mbb R}_{ii}$, where ${\mbb R}_{ii} = 
\int_{L_{i1}}^{L_{i2}} dx_i r (x_i)$. This gives the equation 
$\beta_i - \al_i - (1/K_W) (b_i - a_i) = -(e^2/\pi) (\al_i + \beta_i) 
{\mbb R}_{ii}$, where we have used the fact that $v_W k' = v_F k = 
\om$ and taken the limit $\om \to 0^+$. Using all these equations, 
we recover Eq.~\eqref{cond}. We thus see that the precise profiles 
of $K(x_i)$, $v(x_i)$ and $r(x_i)$ in the different wires are not 
important; all that matters is that the values of $K$ and $v$ are 
given by $K_W$ and $v_W$ as $x_i \to 0+\ep$ and by 1 and $v_F$ as $x_i 
\to \infty$, and that ${\mbb R}_{ii} = \int dx_i r (x_i)$.

\subsection{Conductance for the $M_1$ and $M_2$ classes}

Within the $M_1$ class, the case $\theta =0$ is trivial because $M_1 (0) 
={\mbb 1}$
and $G=0$. We now consider all other values of $\theta$. We find that in
general $G$ depends on $K_W$, $\theta$, and the resistances $R_i =
r_{i0} (L_{i2}-L_{i1})$. [An exception arises for the case $\theta = \pi$
where we find that $G$ is independent of $K_W$ and depends only on the $R_i$.
As shown below, this occurs whenever $M^2 = {\mbb 1}$ which is true for $M_1
(\pi)$ and also for the $M_2$ class for any $\theta$.] The dependence of 
$G$ on $K_W$ for the $M_1$ class is to be contrasted to the case of a single 
wire where the conductance is independent of 
$K_W$~\cite{maslov,pono,safi1,safi2,thomale}. 

In the $M_2$ class, we find that although $G$ depends on $\theta$ and the 
$R_i$, it is completely independent of $K_W$. In Eq.~\eqref{cond}, we write 
${\mbb 1} + M + K_W ({\mbb 1} - M) [{\mbb 1} + (e^2 /\pi) \mbb R] = A + B$, 
where $A = {\mbb 1} + M + K_W ({\mbb 1} - M)$ and $B=(e^2 /\pi) K_W 
({\mbb 1} - M) {\mbb R}$. We can then use the relations that $M^2={\mbb 1}$,
$A^{-1} = [K_W ({\mbb 1}+M)+({\mbb 1}-M)]/ (4K_W)$ and $(A+B)^{-1} = A^{-1} - 
A^{-1} B A^{-1} + A^{-1} B A^{-1} B A^{-1} - \cdots$ to 
show that $G$ does not depend on $K_W$ for any choice of 
$\theta$ and $R_i$. The exact expression for $G$ turns out to be
\bea G &=& - ~\f{e^2}{\pi} ~\f{3 ({\mbb 1}- M_2)}{D}, \nn \\
{\rm where} ~~D &=& 2(\vr_1+\vr_2+\vr_3) + \cos \theta (\vr_1+\vr_2-2\vr_3)
\nn \\
& & -\sqrt{3}\sin \theta (\vr_1-\vr_2), \label{r-inf-m2}\eea
where $\vr_i =1 +(e^2/\pi) R_i$. We thus see that $G$ does not depend on 
$K_W$, the Luttinger parameter in the wire regions.

\section{Power dissipation}
\label{power-dissipation}

In our model, there is no power dissipation exactly at the junction since the 
current splitting matrices $M_1$ and $M_2$ are orthogonal. Power dissipation 
occurs only at the resistive patches and in the leads due to the contact 
resistance. The power dissipation at the contact resistance occurs due 
to the energy relaxation of the electrons in the leads (reservoirs) which are 
maintained at some particular chemical potentials. Classically,
if a current $I_0$ passes through a resistance $R_0$, the power dissipated
$P_0$ is given by $P_0 = I_0^2 R_0$. For a three-wire junction, we can 
define the power dissipated in two equivalent ways as follows
\bsq \begin{align}
P &=& - ~\sum_{i=1}^3 ~V_i ~I_i, ~~~~~~~~~~~~~~~~~~~~~~~~~~~~ \label{P=VI} \\
{\rm and} ~~~P &=& \sum_{i=1}^3 ~I_i^2 ~\Big(R_i+\f{h}{2e^2}\Big).
~~~~~~~~~~~~~~~ \label{P=IsqR} \end{align} \esq
We have verified analytically that these two definitions give the same result.
(A minus sign appears in Eq.~\eqref{P=VI} because we have defined the $I_i$ to
be {\it outgoing} currents).

We know that when the voltages applied in all the three leads are equal to
each other, there should be no current in any of the three wires, and hence 
the power dissipated should be zero. To incorporate this fact, we choose a
new coordinate system for the $V_i$'s known as the Jacobi coordinates
\beq \left[\begin{array}{c}
V_a \\
V_b \\
V_c \end{array}\right] =
\left[\begin{array}{ccc}
1/\sqrt{3} & 1/\sqrt{3} & 1/\sqrt{3}\\
1/\sqrt{2} & -1/\sqrt{2} & 0 \\
1/\sqrt{6} & 1/\sqrt{6} & -2/\sqrt{6}\\
 \end{array}\right] ~\left[\begin{array}{c}
V_1 \\
V_2 \\
V_3 \end{array}\right]. \label{Jacobi-V} \eeq

In this coordinate system, the power dissipated $P$ does not depend on 
the voltage $V_a$. Further, it turns out that $V_b$ and $V_c$ can be 
parameterized in such a way that $P$ depends only on one of two parameters; 
this parametrization is different for junctions described by $M_1$ and $M_2$ 
as we will see below. Since we know from Eq.~\eqref{P=IsqR} that the power 
dissipated at the resistance $R_i$ is $I_i^2 R_i$, for simplicity we will 
only consider the case of $R_i=0$ (for $i=1,2,3$) for further analysis.

\subsection{Power dissipated for the $M_1$ class}

\begin{figure}[htb]
\begin{center}
\epsfig{figure=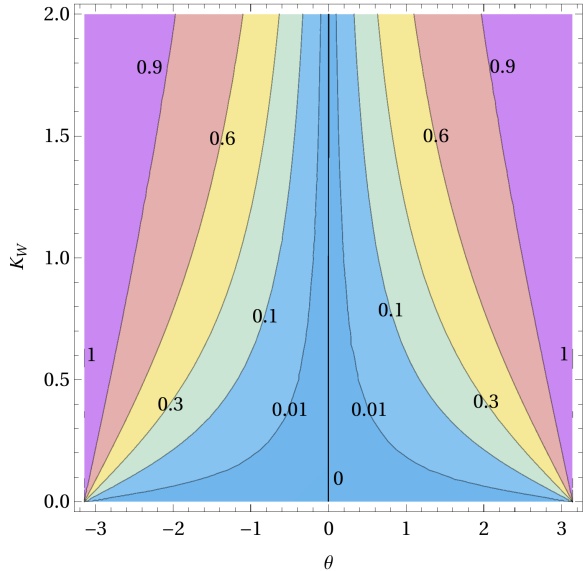,width=8cm}
\end{center}
\caption{Power dissipated for a $Y$-junction (see Fig.~\ref{Y-jn-diagram})
shown as a contour plot in the $\theta - K_W$ plane for the $M_1$ class. The 
numbers shown indicate the power dissipated along the nearest contour in units
of $e^2/h$; we have set $V_0=1$ and $R_i = 0$.} \label{M1-Kw} \end{figure}

In this case, the $M$-matrix is invariant under a cyclic 
permutation of the wires 1, 2 and 3, and the dissipated power $P$ 
turns out to be invariant under rotations in the
$V_b-V_c$ plane. If we write $V_b=V_0\cos\phi$ and 
$V_b=V_0\sin\phi$, the power dissipated is proportional to $V_0^2$ and 
does not depend on $\phi$. However, $P$ depends on the parameters $K_W$ 
and $\theta$ since the conductance matrix $G$ depends on 
those two parameters. The dependence of $P$ on $K_W$ and 
$\theta$ is shown as a contour plot in Fig.~\ref{M1-Kw}. The choice 
$\theta=0$ decouples the three wires at the junction, making the currents 
in the three wires zero; hence, the power dissipated is zero for this choice. 
The power dissipated is maximum on the contour $\theta=\pm\pi$.

\subsection{Power dissipated for the $M_2$ class}

Junctions described by $M_2$-class are time reversal invariant
(the $M$-matrix is symmetric).
In this case, the power dissipated is a constant over straight lines in the 
$V_b-V_c$ plane. If we define the variables $V_{\pm} = V_b \cos (\theta/2) 
\pm V_c \sin (\theta/2)$, then the power dissipated $P$ 
is found to be independent of $V_+$ and is given by $P= (e^2/h) ~V_{-}^2$.
Further, $P$ does not depend on $K_W$ in this case.

\section{Resistances in parallel}
\label{resistance-parallel}

At the end of Sec.~\ref{single-wire} we saw that the effective 
resistance of two or more resistances in series is the sum of the 
individual resistances in agreement with the classical result. 
By the word `classical' we mean the result obtained for the effective
resistance by using Kirchoff's circuit laws. In this spirit, it is
interesting to study whether the effective resistance of two resistances 
in parallel will agree with the classical result for a similar system. 

\begin{figure}[htb]
\epsfig{figure=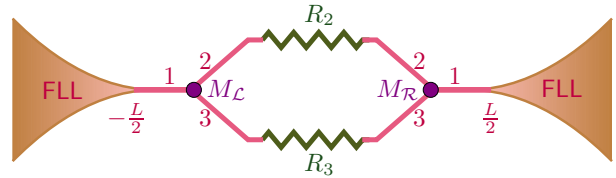,width=8cm}
\caption{Schematic diagram of resistances $R_2$ and $R_3$ in parallel 
attached to Fermi liquid leads through junctions ${\mc L}$ and ${\mc R}$.} 
\label{parallel_res} \end{figure}

To begin with, let us consider a general model illustrated in 
Fig.~\ref{parallel_res}. The three wires at each junction are labeled by the 
index $i=1,2,3$, and the two junctions are described by current splitting 
matrices $M_{\mc L}$ and $M_{\mc R}$. The coordinate $x_i$ runs from
\newline (i) ~~$-\infty$ to $-l$ on the wire $i=1$ on the left,
\newline (ii) ~$-l$ to $l$ on the wires $i=2, ~3$,
\newline (iii) $l$ to $\infty$ on the wire $i=1$ on the right.
\newline The resistive regions on the wires $i=2,3$ lie in the range $|x_i|<a$,
the total resistances being $R_i$. In the leads, $K=1$ and $v=v_F$ as usual,
while $K=K_W$ and $v=v_W$ in the regions $a<|x_i|<L/2$ shown in pink
(solid lines). In the 
resistive regions, $K(x)$ and $v(x)$ can have any profiles and these do not 
affect the final result. The incoming and outgoing fields at the two junctions 
(at $x_i = \pm l$) are related by the matrices $M_{\mc R}$ and $M_{\mc L}$. 
A plane wave incident from the left has the solution $\phi_{i,k}(x_i,t) 
= f_{i,k}~e^{-i\om t}$ given by: 
\bsq \begin{align} 
f_{1,k} &=& e^{ikx}+s_k e^{-ikx}~~~~~~~~~~~~~~~~~~{\rm~~for~~}x_1<-L/2,\nn \\
&=& t_{1{\mc L},k}~e^{ikx} + s_{1{\mc L},k}~e^{-ikx} {\rm~~for~}-L/2<x_1<-l,
\label{par-sol-1L} \\
f_{j,k} &=& t_{j{\mc L},k}~e^{ikx} + s_{j{\mc L},k}~e^{-ikx} ~~~{\rm~~~for~}
-l<x_j<-a, \nn \\
&=& t_{j{\mc R},k}~e^{ikx} + s_{j{\mc R},k}~e^{-ikx} ~~~~~~~~~{\rm~~~for~~}
a<x_j<l \nn \\
&&{\rm for~}j=2 ~~{\rm and}~~3, ~~~~~~~~~~~~~~~~~~~~~~~~~~~~~~~~~~~~~~~
\label{par-sol-23} \\
f_{1,k} &=& t_{1{\mc R},k}~e^{ikx} + s_{1{\mc R},k}~e^{-ikx} ~~~~~~
{\rm~~for~~} l<x_1< L/2, \nn \\
&=& t_k e^{ikx}~~~~~~~~~~~~~~~~~~~~~~~~~~~~~~~~{\rm~~for~~}x_1>L/2.
\label{par-sol-1R} \end{align} \esq
A calculation similar to the one performed in Sec.~\ref{single-wire-scatt-app} 
relates (in the DC limit) the unknowns $t_{j{\mc P},k}$ and $s_{j{\mc P},k}$ 
(for {\small{$\mc{P = L,R}$}} and $j=2,3$) as
\bea \left[\begin{array}{c}
t_{j\mc{L}k} \\
s_{j\mc{L}k} \end{array}\right] &=&
\left[\begin{array}{cc}
1 + \vr_j & \vr_j \\
-\vr_j & 1-\vr_j \end{array}\right] ~\left[ \begin{array}{c}
t_{j\mc{R}k} \\
s_{j\mc{R}k} \end{array}\right], \label{t-LR} \eea
where $\vr_j = e^2 R_j/(2\pi)$. The amplitudes $t_{j{\mc P},k}$ and 
$s_{j{\mc P},k}$ for different $j$'s are related by the matrix $M_{\mc P}$
as follows: 
\bea w^{(out)}_{\mc P} &=& M_{\mc P} ~w^{(in)}_{\mc P} ~~~{\rm~for~~}
\mc{P=L,R}, \nn \\
{\rm where} ~~w^{(out)}_{\mc L} &=& [-s_{1\mc{L},k}~~~~t_{2\mc{L},k}~~~~
t_{3\mc{L},k}]^T, \nn \\
w^{(in)~}_{\mc{L}} &=& [~t_{1\mc{L},k}~~-s_{2\mc{L},k}~~-s_{3\mc{L},k}]^T, 
\nn \\
w^{(out)}_{\mc{R}} &=& [~t_{1\mc{R},k}~~-s_{2\mc{R},k}~~-s_{3\mc{R},k}]^T, 
\nn \\
w^{(in)~}_{\mc{R}} &=& [-s_{1\mc{R},k}~~~t_{2\mc{R},k}~~~t_{3\mc{R},k}]^T.
\label{M-P} \eea
This is essentially same as the relation $b=-M ~a$ used in 
Sec.~\ref{junc-scatt}. The relations \eqref{t-LR} and \eqref{M-P} along with 
the continuity of $\phi_{1,k}$ and $\f{v}{K} \Do_x \phi_{1,k}$ at $x=\pm L/2$ 
give us enough conditions to solve for all the unknowns, namely, $s_k$, 
$t_{j{\mc P},k}$, $s_{j{\mc P},k}$ and $t_k$.

Once we know the final expression for the DC conductance $\si_{dc}$ ($= 
\f{e^2}{2\pi}~ t_k$ in the limit $k \to 0^+$), we can obtain the effective 
resistance $R_{||}$ by subtracting out the contact resistance from the total 
resistance, 
\beq R_{||} ~=~ \f{1}{\si_{dc}} - \f{h}{e^2}. \label{eff-res} \eeq
We have listed the results obtained for different choices of $M_{\mc L}$
and $M_{\mc R}$ in Table~\ref{tab.ML-MR}.

\begin{table}[htb]
\begin{center}
\begin{tabular}{|c|c|c|}
\hline
$M_{\mc L}(\theta_{\mc L})$ & $M_{\mc R}(\theta_{\mc R})$ & Expression for 
$R_{||}$ \\
\hline
$M_1(\theta)$ & $M_1(-\theta)$ & $R_2R_3/(R_2+R_3)$ \\
$\theta \ne 0$ & & \\
\hline
$M_1(\theta_{\mc L})$ & $M_1(\theta_{\mc R})$ & Depends on $\theta_{\mc L}$,
$\theta_{\mc R}$ and $K_W$ \\
$\theta_{\mc L}\ne -\theta_{\mc R}$~ & & as shown in Figs. \ref{M1-M1-Kw} and 
\ref{M1-M1-th} \\
\hline
$M_1(\theta_{\mc L} = 0)$ & $M_2(\theta_{\mc R})$ & $\infty$ \\
\hline
$M_1(\theta_{\mc L} )$ & $M_2(\theta_{\mc R})$ & Depends only on 
$\theta_{\mc R} =\theta$ \\
& & as shown in Fig. \ref{M1M2_n_M2M2-th} \\
\hline
$M_2(\theta)$ & $M_2(\theta)$ & Depends only on $\theta$ \\ & & as shown in 
Fig. \ref{M1M2_n_M2M2-th} \\
\hline
$M_2(\theta_{\mc L} \ne \theta_{\mc R})$ & $M_2(\theta_{\mc R})$ & $\infty$ \\
\hline
\end{tabular}
\end{center}
\caption{The behavior of the effective resistance $R_{||}$ for different 
choices of $M_{\mc L}$ and $M_{\mc R}$.} \label{tab.ML-MR} \end{table}

\begin{figure}[htb]
\begin{center}
\epsfig{figure=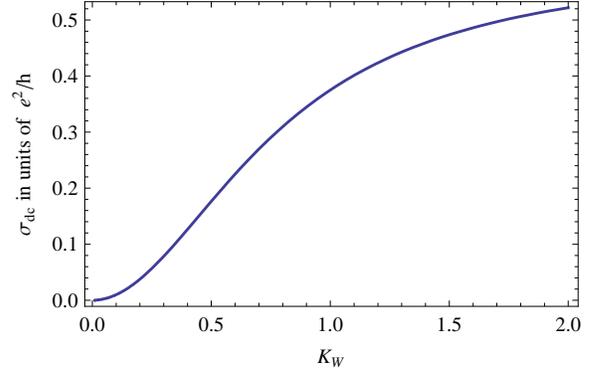,width=8cm}
\end{center}
\caption{DC conductance is plotted versus $K_W$ for the case that $M_{\mc L}$ 
and $M_{\mc R}$ lie in the $M_1$ class. Parameters: $\theta_{\mc L} =
\theta_{\mc R} =\pi/2$, $\vr_2=1$ and $\vr_3=2$.} \label{M1-M1-Kw} \end{figure}

\begin{figure}[htb]
\begin{center}
\epsfig{figure=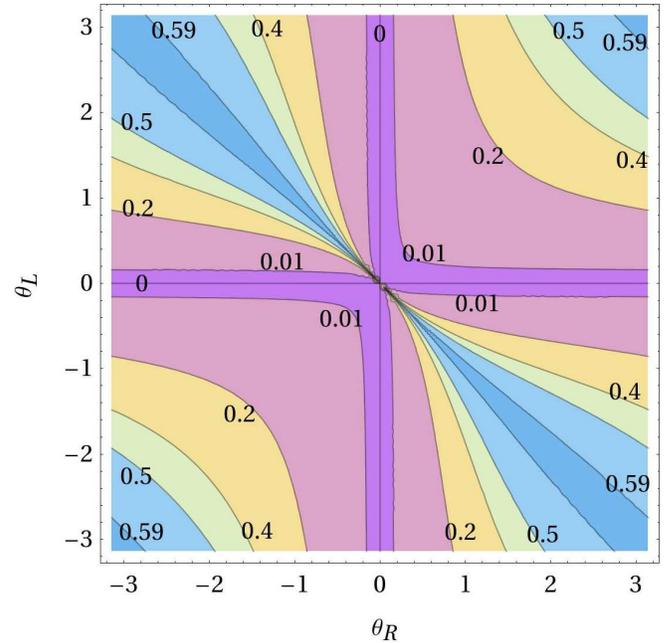,width=8.6cm}
\end{center}
\caption{DC conductance $\si_{dc}$ for a parallel combination of resistances 
(see Fig.~\ref{parallel_res}) shown as a contour plot in the
$\theta_{\mc R}-\theta_{\mc L}$ plane for a fixed interaction strength 
$K_W=0.6$, when both $M_{\mc L}$ and $M_{\mc R}$ lie in the $M_1$ class. 
We have chosen 
$\vr_2=1,~\vr_3=2$. The numbers indicate the values of $\si_{dc}$ 
on the contours in units of $e^2/h$.} \label{M1-M1-th} \end{figure}

In an earlier paper, we observed that in the case of a three-wire junction, 
if time reversal invariance is broken at the junction (i.e., if $M$ is not a 
symmetric matrix), then the conductance matrix depends on $K_W$~\cite{soori}.
In the model studied here, we find that the final DC conductance depends on 
$K_W$ only when time reversal invariance is broken at both the junctions
(i.e., if both $M_{\mc L}$ and $M_{\mc R}$ belong to the $M_1$ class) and 
$\theta_{\mc L}\ne-\theta_{\mc R}$. The dependence of the conductance 
$\si_{dc}$ on $K_W$ ($\theta_{{\mc L}/{\mc R}}$) is shown in 
Fig.~\ref{M1-M1-Kw}(\ref{M1-M1-th}). The contour $\theta_{\mc L}
=-\theta_{\mc R}=\theta \ne 0$ corresponds to the maximum value of
 $\si_{dc}$. On this contour, $\si_{dc}$ does not depend on either $K_W$ 
or $\theta$; moreover we get an expression for the effective resistance 
$R_{||}$ which agrees with the classical result for $R_{||}$.

\begin{figure}[htb]
\begin{center}
\epsfig{figure=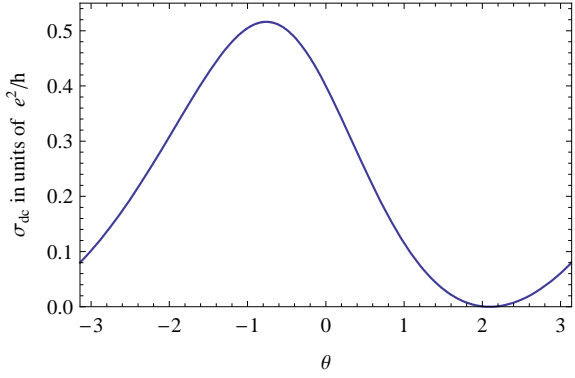,width=8cm}
\end{center}
\caption{DC conductance is plotted versus $\theta$ for the cases
(i) $M_{\mc L}$ and $M_{\mc R}$ lie in the $M_2$ class and $\theta_{\mc L}=
\theta_{\mc R}= \theta$, and
(ii) $M_{\mc L}$ lies in the $M_1$ class and $M_{\mc R}$ lies in the $M_2$
class, $0< \theta_{\mc L} < 2\pi$ and $\theta_{\mc R} = \theta$. Parameters:
$\vr_2=1$ and $\vr_3=2$.} \label{M1M2_n_M2M2-th} \end{figure}

The choice $\theta=0$ at a junction described by $M_1$-matrix decouples all the
three wires at the junction. Hence, the conductance of the system is zero (or 
equivalently $R_{||}=\infty$) if either of the two junctions have $\theta=0$.
We can see this in both Table~\ref{tab.ML-MR} and 
Fig.~\ref{M1-M1-th}. Another interesting case that results in infinite 
effective resistance arises when both $M_{\mc L}$ and $M_{\mc R}$ lie
in the $M_2$ class and $\theta_{\mc L}\ne\theta_{\mc R}$.
The choices, (i) $M_{\mc L}$ and $M_{\mc R}$ lie in the $M_2$ class with
$\theta_{\mc L}~=~\theta_{\mc R}~=~\theta$ and (ii) $M_{\mc L}$ lies in the
$M_1$ class and $M_{\mc R}$ lies in the $M_2$ class, with $0< \theta_{\mc L}
< 2\pi$ and 
$\theta_{\mc R}=\theta$, yield the same $\theta$-dependent expression for the 
conductance, as shown in Fig.~\ref{M1M2_n_M2M2-th} for one particular choice 
of the parameters. The case $M_{\mc L}(\theta_{\mc L}) = M_2(\theta_2)$ and
$M_{\mc L}(\theta_{\mc R}) = M_1(\theta_1)$ is redundant since we have 
already analyzed the case $M_{\mc L}(\theta_{\mc L}) = M_1(\theta_1)$ and 
$M_{\mc L}(\theta_{\mc R}) = M_2(\theta_2)$, and these two cases 
are equivalent. The equivalence of these two cases follows from the
parity transformation $x_i\to -x_i$, $i=1,2,3$, since the conductance is 
parity invariant for this system.

Finally, as a special case, we look at a symmetric situation treating the 
wires 2 and 3 on the same footing, with both junctions being described by the 
same $M$-matrix. The case $M=M_1$ requires $\theta = \pi$ for symmetry 
between the wires 2 and 3. As $\theta_{\mc L} = - \theta_{\mc R}$ for
$\theta_{\mc L} = \pi$, we conclude from Table~\ref{tab.ML-MR} that $R_{||}=R_2
R_3/(R_2+R_3)$. Turning to the case $M=M_2$, the symmetry requirement 
imposes $\theta=-\pi/3$. The effective resistance in this case takes the 
form $R_{||}=(R_2+R_3)/4$. An interesting implication of this is 
that when one of the resistances (say $R_3$) is taken to infinity keeping 
the other ($R_2$) finite, the effective resistance $R_{||}$ also goes to 
infinity. [This is unlike the classical case where, if one of the 
resistances ($R_3$) in a parallel geometry is taken to infinity, the 
effective resistance approaches the value of the other resistance ($R_2$)].
 We will provide an understanding of this surprising result in 
Sec.~\ref{pulse} by a physically intuitive argument.

\section{Transmission of a pulse}\label{pulse}

Inspired by the scattering method~\cite{soori,safi1,safi2}, we illustrate 
a `sum over paths' method to calculate the DC conductance of a system. 
We consider a special case of the model studied in 
Sec.~\ref{resistance-parallel} by setting $K=1$ everywhere and $R_3=\infty$. 
We now allow a $\de$-pulse to be incident from the left lead and calculate
the transmission amplitude for the pulse to exit at the right lead.
(In the bosonic language, the conductance is just proportional to the 
transmission amplitude). We know that the pulse broadens when reflected from 
a resistive region of finite width and that the width of the reflected pulse 
is twice the width of the resistive region~\cite{soori}. For simplicity, let
us choose the pulse to be a $\de$-function. In order to maintain the pulse 
as a $\de$-function even after scattering from the resistive region,
we choose the resistivity profile for $R_2$ to also be a 
$\de$-function given by $r_2(x_2)=R_2 ~\de(x_2)$. We have seen earlier in 
Sec.~\ref{single-wire} that doing this does not alter the final expression for
the conductance as long as the total resistance of the patch remains the same.

\begin{figure}[htb]
\epsfig{figure=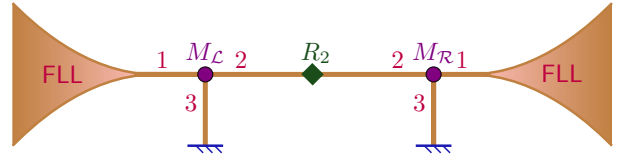,width=8cm}
\caption{Schematic diagram of the model. The $\de$-function resistivity 
profile is highlighted by the diamond in the middle. Any pulse which hits 
the free end of either of the wires labeled 3 gets completely reflected.} 
\label{two-stub} \end{figure}

\begin{figure}[htb]
\epsfig{figure=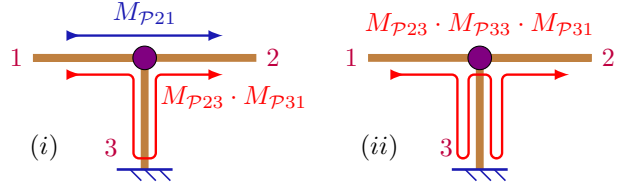,width=8cm}
\caption{Transmission amplitudes for a few paths are shown as examples.} 
\label{stub-SOP} \end{figure}

Taking the limit $R_3\to\infty$ is like attaching stubs on either sides of
the resistance $R_2$ and connecting it to Fermi liquid leads as shown in 
Fig.~\ref{two-stub}. To illustrate the `sum over paths' method, let us 
first consider a simpler model where we 
calculate the transmission amplitude at a $Y$-junction formed by attaching a 
single stub to a wire as shown in Fig.~\ref{stub-SOP}. Let $M_{\mc P}$ be the 
$M$-matrix that relates the outgoing bosonic fields to the incoming bosonic 
fields at the junction. Of the pulse approaching the $Y$-junction through the 
wire $i$, a fraction $M_{\mc P ji}$ goes into the wire $j$ after scattering.
So the transmission amplitude for the path $1\to 2$ is $M_{\mc P 21}$. Since 
the pulse gets completely reflected at the free end of wire 3, the 
transmission amplitude for the path $1 \to 3 \to 2$ is $M_{\mc P 23} ~
M_{\mc P 31}$. 

The total transmission amplitude for this system is a sum over transmission 
amplitudes for all possible paths. In addition to the two paths shown in 
Fig.~\ref{stub-SOP} (i), there are infinitely many paths in which the pulse 
starting in wire 1 ends in wire 2. These paths are characterized by 
multiple reflections of the pulse between the junction-$\mc P$ and the free 
end of the wire 3. One such path is shown in Fig.~\ref{stub-SOP} (ii). The 
path $1\to~(3\to~3)^n~\to~2$ from wire 1 to wire 2 with $n$ reflections 
between junction-$\mc P$ and the free end of the wire 3, has a transmission 
amplitude of $M_{\mc P 23} ~(M_{\mc P 33})^n ~M_{\mc P 31}$ (this is non-zero
only when $\theta_{\mc P}\ne0$). Hence the total transmission amplitude 
$t_{\mc P}(1\to2)$ (when $\theta_{\mc P}\ne0$) turns out to be
\beq t_{\mc P}(1\to2)~=~M_{\mc P 21}+\f{M_{\mc P 23} 
M_{\mc P 31}}{1-M_{\mc P 33}}. \label{t-12} \eeq
The DC conductance of such a system connected to leads on the two sides, 
is given by $\si_{dc}~=~ (e^2/h)~ t_{\mc P}(1\to2)$. Using the fact that the 
matrix $M_{\mc P}$ describing the junction is parameterized by a single 
variable $\theta_{\mc P}$, one can show that (i)~$\si_{dc}=e^2/h$ for $M_{\mc 
P}$ lying in the $M_1$ class, and 
(ii)~$\si_{dc}=0$ for $M_{\mc P}$ lying in the $M_2$ class, irrespective of 
the angle $\theta_{\mc P}$ as long as $\theta_{\mc P} \ne 0$. For the $M_1$ 
class, $\theta_{\mc P} = 0$ means that the three wires are disconnected and 
it is easy to see that $\si_{dc}=0$, while for the $M_2$ class, $\theta_{\mc P}
= 0$ means that the stub is disconnected from the wire and the calculation 
gives the expected result, $\si_{dc}=e^2/h$.

It is easy to see by this method that the effective resistance of a system 
having two or more scatterers in series is the sum of the individual 
resistances. In other words, if $t_1$, $t_2$, ... are the transmission 
probabilities of the different scatterers in series, the quantities
$(1/t_i - 1)$'s add up 
to give $1/t - 1$, where $t$ is the transmission probability for the system 
(see pp.~63-64 of Ref.~\cite{datta}). Now, the model in Fig.~\ref{two-stub} 
can be looked upon as a system with three scattering centers, i.e., a
resistance $R_2$ in the middle and two $Y$-junctions on the two sides of 
$R_2$. When either of the two junctions ($\mc P$) is described by an 
$M_2$-matrix (with $\theta\ne0$), we get $R_{||}=\infty$ since the resistance
of the scatterer at $\mc P$ is infinite. When both the junctions are described
by $M_1$-matrix (with the constraints that $\theta_{\mc L}\ne0$ and 
$\theta_{\mc R} \ne0$), we get $R_{||}=R_2$.

\section{Spin-1/2 electrons}
\label{spin-12}

Let us briefly discuss an extension of our results to the realistic case of 
spin-1/2 electrons. In one dimension, it is known that interactions between 
electrons lead to the phenomenon of spin-charge separation~\cite{boson}. A 
bosonic description of the system begins with fields for spin-up and spin-down
electrons denoted by $\phi_\ua$ and $\phi_\da$ respectively. The fields for 
the charge and spin modes are then given by
\bea \phi_c &=& \f{1}{\sqrt 2} ~(\phi_\ua ~+~ \phi_\da), \nn \\
{\rm and} ~~~\phi_s &=& \f{1}{\sqrt 2} ~(\phi_\ua ~-~ \phi_\da). \eea
The system decouples in terms of these fields even when density-density 
interactions are introduced between the electrons. The Lagrangian is
similar in form to the one in Eq.~(\ref{lag1}) except that there are
two sets of parameters denoted by $(K_c,v_c)$ and $(K_s,v_s)$ for 
the charge and spin fields respectively. Namely,
\bea L &=& \int_{-\infty}^\infty dx ~[\f{1}{2v_cK_c} (\Do_t \phi_c)^2 ~-~
\f{v_c}{2K_c} (\Do_x \phi_c)^2 \nn \\
& & ~~~~~~~~~+~ \f{1}{2v_sK_s} (\Do_t \phi_s)^2 ~-~ \f{v_s}{2K_s} 
(\Do_x \phi_s)^2],
\label{lag2} \eea
where we have ignored a term involving the cosine of the field $\phi_s$ 
arising from 
fourth-order fermionic terms like $(\psi_\ua^\dg \psi_\da)^2$~\cite{boson}. 
For an inhomogeneous system, the parameters $K_a$ and $v_a$ are generally
functions of $x$. However, for a system with $SU(2)$ rotational invariance, 
$K_s = 1$.

The charge and spin currents are given by
\bea j_c &=& \f{e}{\sqrt \pi} ~{\sqrt 2} ~\Do_t \phi_c, \nn \\
{\rm and} ~~~j_s &=& \f{e}{\sqrt \pi} ~{\sqrt 2} ~\Do_t \phi_s \label{jcs} \eea
respectively. We note that $j_c$ is invariant under $SU(2)$ rotations,
while $j_s$ is only invariant under $U(1)$ rotations about the $z$-axis.
The simplest way of introducing resistance in this theory would be to 
postulate a Rayleigh dissipation function for the charge current given by 
\beq {\cal F} ~=~ \f{1}{2} ~\int_{-\infty}^\infty dx ~r ~j_c^2 ~=~
\f{e^2}{2\pi} ~\int_{-\infty}^\infty dx ~2r ~(\Do_t \phi_c)^2. 
\label{diss2} \eeq
We can now use this dissipative function to calculate the charge conductance 
in a variety of systems as in the spinless case. In principle, one can also 
introduce a dissipative function for the spin current; however such a function
would have both a term quadratic in the field $\phi_s$ as well as a cosine of
$\phi_s$~\cite{rech}. This makes it difficult to analyze the corresponding
equations of motion.

In analogy with Eqs.~(\ref{sigma-G}-\ref{G}), we can compute the two-point
correlation function of the charge current $j_c$ to find the charge 
conductance $G_c$ for a three-wire junction. The presence of the factors 
of $\sqrt 2$ in Eq.~\eqref{jcs} and 2 in Eq.~\eqref{diss2} as compared 
to the corresponding expressions for the spinless case implies that $G_c$
will be given by an expression similar to Eq.~\eqref{cond}, except for some
factors of 2. Namely, we will have
\bea G_c &=& - \f{2e^2 K_{cW}}{\pi} ~[ {\mbb 1} + M + K_{cW} ({\mbb 1} - M) 
({\mbb 1} + \f{2e^2}{\pi} \mbb R)]^{-1} \nn \\
& & ~~~~~~~~~~~\times ~[ {\mbb 1} - M], \label{cond4} \eea
where the current splitting matrix $M$ and the resistance matrix $\mbb R$
are defined as before, and $K_{cW}$ denotes the Luttinger parameter for the
charge field in the wire regions given by $0 < x_i < L_{i1}$.

One can actually define two conductances, $G_c$ and $G_s$, which 
govern the amounts of charge and spin currents which flow when the 
corresponding voltage biases or 
chemical potential differences are applied between different leads,
namely, $e \De V_c = \De \mu_\ua = \De \mu_\da$ and $e \De V_s
= \De \mu_\ua = - \De \mu_\da$ for driving charge and spin 
currents respectively. 
In the absence of resistances, it 
has been shown in Ref.~\cite{hou} that for a junction of three quantum wires,
the RG flows resulting from the interactions between the electrons generally 
take the system to a fixed point at which the charge and spin conductances, 
$G_{0c}$ and $G_{0s}$, are different from each other. [We note here that 
the RG flows occur entirely within the interacting regions $0 \le x_i \le 
L_{i1}$ in Fig.~\ref{Y-jn-diagram}, and that no renormalization occurs in 
the dissipative regions since there is no interaction in those regions]. 

Let us now introduce resistances in the system. If we then use a relation 
like Eq.~\eqref{cond3} to find $G_a$ from $G_{0a}$ and $\mbb R$, where 
$a=c,s$, we expect to find that $G_c$ and $G_s$ would also not be equal 
to each other since $G_{0c}$ and $G_{0s}$ differ from each other. Note that
this is purely an effect of interactions between the electrons; for
non-interacting electrons, we would have $G_{0c} = G_{0s}$ and therefore
$G_c = G_s$.

\section{Discussion}

We have developed a formalism which allows us to study the effect of 
resistances in a quantum wire using the technique of bosonization. The 
analysis can be extended to a system of three wires by using a current 
splitting matrix $M$ to describe the junction. It is known that there are 
two classes of such matrices which are called $M_1$ and $M_2$. We have 
calculated the conductance of a three-wire system using both Green's function
and scattering approaches. We have then examined the power dissipated by the 
system as a function of the matrix $M$ and the voltages applied in the three 
leads. For the $M_1$ class, both the conductance and the dissipated power 
depend on the value of the interaction parameter $K$ near the junction, 
while in the $M_2$ class, the 
conductance and power are independent of $K$. Next, by putting together 
two junctions of three wires, we have studied the effective conductance 
of a system of two wires in parallel. This is found to depend in a highly
non-trivial way on the $M$ matrices at the two junctions, the parameter
$K$, and the resistances in the two wires. In some cases, the effective 
resistance is infinite, while in other cases, it is finite but depends on $K$.
In only one special case do we obtain the classical result for two parallel 
resistances. For the case in which one of the two resistances is infinitely 
large, we have provided an intuitive way of calculating the effective 
conductance by summing over all the paths that an electron 
can take in going from one lead to the other. This method also shows that
three-wire junctions with matrices $M_1$ and $M_2$ behave quite differently
from each other when the resistance in one of the wires is taken to be
infinitely large. Finally, we have generalized our results to the case of 
interacting spin-1/2 electrons. We have argued that the charge and spin 
conductances will generally be different from each other due to RG flows 
induced by interactions between the electrons.

The formalism discussed in this paper is well suited for studying systems with
interacting electrons in which the resistances are phase incoherent. It would
be useful to develop a more general method which can deal with partially 
coherent resistances. It would also be useful to have a more complete 
treatment of systems with spin-1/2 electrons in which the charge and spin 
resistivities are position dependent and different from each other, in the 
spirit of Ref.~\cite{rech}.

\section*{Acknowledgments}
We thank Sourin Das for stimulating discussions. A. S. thanks Abhishek R. Bhat
for help with the numerics. A. S. thanks CSIR, India for financial support, 
and D. S. thanks DST, India for financial support under SR/S2/JCB-44/2010.

\end{document}